\documentclass[11pt,twoside]{article}

\usepackage{asp2004}
\usepackage{epsf}
\usepackage{graphicx}
\usepackage{psfig}
\usepackage{lscape}

\begin{document}
\title{Mid-Infrared Spectroscopic Diagnostics of Galactic Nuclei}
\author{Vassilis Charmandaris \& the Spitzer/IRS Instrument Team}
\affil{University of Crete, Greece}

\begin{abstract} 
In this paper I summarize the science motivations, as well as a few
mid-infrared spectroscopic methods used to identify the principal
mechanisms of energy production in dust enshrouded galactic
nuclei. The development of the various techniques is briefly
discussed.  Emphasis is given to the use of the data which are
becoming available with the infrared spectrograph (IRS\footnote{The
IRS was a collaborative venture between Cornell University and Ball
Aerospace Corporation funded by NASA through the Jet Propulsion
Laboratory and the Ames Research Center.}) on Spitzer, as well as the
results which have been obtained by IRS over the past two years.
\end{abstract}

\vspace*{-0.5cm}
\section{Introduction}
\vspace*{-0.3cm}

One of the open issues in extragalactic astronomy is how to quantify
the physical mechanisms that contribute to the energy production in
galactic nuclei. The most luminous galaxies often display strong
evidence of massive star formation taking place in their nucleus. This
is typically due to the large quantities of atomic and molecular gas
driven in their center as a result of instabilities on their disk such
as bars, which often form during tidal interactions. It is also widely
accepted that most galactic nuclei likely harbor a super-massive black
hole (SMBH) \citep[see review of][]{Ferrarese05}. Due to the high
densities found in the nuclei the SMBH may accrete matter at a
variable rate, even though this accretion does not always result in
emission of radiation \citep{Narayan98}, which would characterize the
galaxy as harboring an active nucleus (AGN). A prime example of such
``non-AGN'' is our own Galactic Center, which does contain a
$\sim$10$^6$M$_\odot$ SMBH and accretes material, but does not display
any visible AGN activity \citep{Genzel87, Ghez98}. However, in the
cases where AGN activity is present, hard electromagnetic radiation,
mostly in form of UV and X-rays, is emitted from the nucleus. This is
typically observed either directly, in the form of X-rays and radio
emission which are less affected by obscuration, and/or indirectly as
high energy photons are absorbed by the gas and dust and subsequently
re-emitted at longer wavelengths. So the original question naturally
translates to what fraction of the bolometric energy observed from a
galaxy originates from an accretion disk (AGN) compared to usual star
formation activity. This issue is of particular interest in cases
where most of the energy of a galaxy originates from the nuclear
regions rather than the outer regions/disk.

Historically, our knowledge on the physics of the nuclear activity of
galaxies is mostly based on optical and near-IR spectroscopy, obtained
in major ground based telescope facilities. A wealth of diagnostic
methods to classify and quantify the AGN activity have been developed
over the years \citep[ie][]{Veilleux87, Armus89, Kennicutt98,
Kewley01}. A presentation of these methods is well beyond the scope of
this review but it is only fair to state that they have been extremely
successful in providing insight into the properties of galactic
nuclei, in particular in cases where a direct line of sight to the
nucleus is available. Polarization measurements have also been very
useful in identifying evidence of AGN activity, in cases where the
nucleus is indirectly probed via scattered light \citep{Heisler97}.

The effect of dust absorption though, which does vary substantially as
a function of wavelength, is one of the principal reasons which
complicate the interpretation of the optical data. It is useful to
remind ourselves that a 10eV optical photon can penetrate $\sim$0.5mag
of dust, while a 1keV X-ray photon can pass through a hydrogen column
of $\sim10^{22}$cm$^{-2}$. If we were to consider the case of our
Galactic Center where the A$_{\rm V}\sim$30mag, only $\sim$1 in
10$^{12}$ optical photons emitted in the nucleus of the Milky Way can
reach our Sun. However, if we were to observe in K-band (2.2$\mu$m)
the A$_{\rm 2.2\mu m}\sim$2.5mag, and one can detect 10\% of the
near-IR photons emitted from the source.

Based on the arguments above, X-rays are clearly the best diagnostic
of an AGN especially for higher redshift systems, as the slope of the
X-ray spectrum and the sensitivity of the detectors result in
relatively flat sensitivity out to z$\sim$4--5
\citep{Brandt05}. However, the energy emitted in X-rays from the
majority of luminous extragalactic sources is
$\sim10^{42}$erg\,s$^{-1}$, several orders of magnitude bellow their
bolometric luminosity. This, in addition to the fact that X-ray
observations need to be performed from space, and most extragalactic
sources are intrinsically faint in X-rays, makes estimates of the
contribution of an AGN to their bolometric luminosity rather
challenging \citep[see][]{Mushotzky04}. Radio waves are an appealing
alternative, in particular given the latest improvements in VLBA
\citep{Lonsdale93,Lonsdale03}.  Even in this case though, there are
difficulties in detecting higher redshift systems, as well as in
disentangling issues related to non-thermal emission and
self-absorption from embedded nuclei \citep[see][]{Condon92}.

The use of the infrared part of the spectrum $3<\lambda<1000\mu$m in
order to develop AGN diagnostics is an avenue which has been explored
for the past twenty years for two main reasons: a) most luminous
galaxies, QSOs being a ``bright'' exception, emit a sizeable fraction
their energy ($>$30\% in the infrared and b) the infrared is
considerably less affected by dust absorption than the optical.

The all sky survey by IRAS provided the first opportunity to classify
a large number of galaxies based on the shape of their global spectral
energy distribution (SED). \citet{deGrijp87} showed that a ratio of
S60$/$S25$>$0.26\footnote{Where S25 and S60 are the 25 and to 60
$\mu$m IRAS flux densities.} could result in a 70\% success rate in
identifying previously unknown Seyferts. More recent developments on
broadband mid-IR diagnostics are presented in another review on this
volume.

\vspace*{-0.3cm}  
\section{Mid-IR Emission Line Diagnostics}
\vspace*{-0.3cm}

\begin{figure}[!h]
%\epsscale{0.8}
%\plotone{figure1.eps}
\centerline{\includegraphics[scale=0.55]{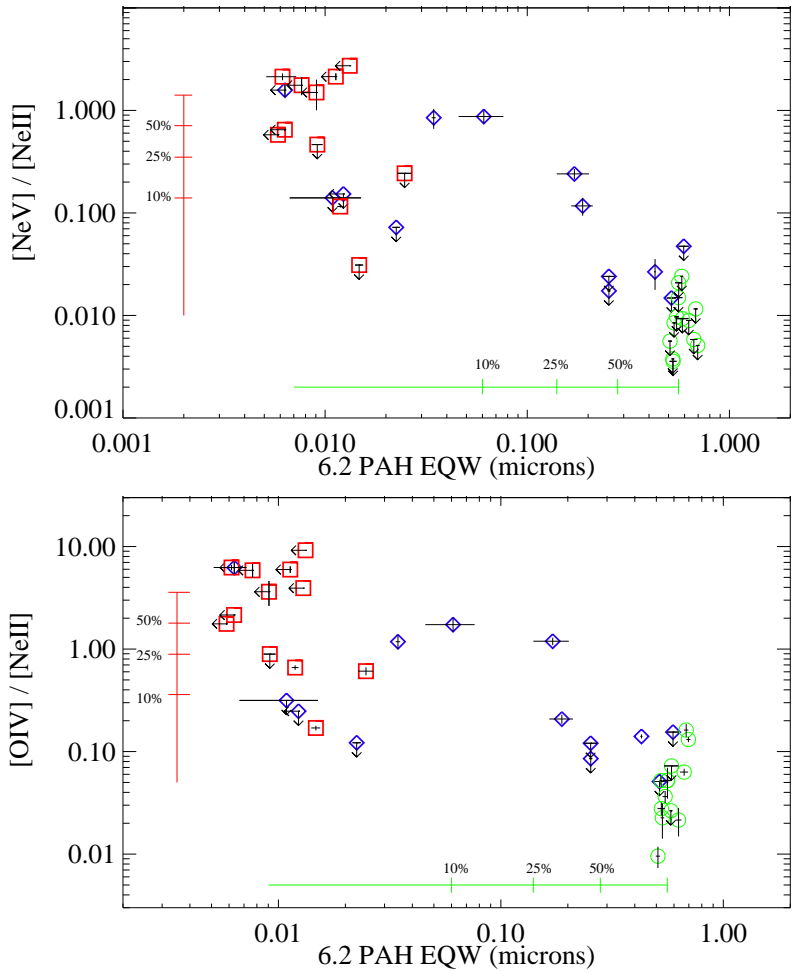}}
\vspace*{-0.2cm}
\caption{Mid-IR excitation diagrams for [NeV]$\lambda14.3\mu$m (top)
  and [OIV]$\lambda25.9\mu$m (bottom) as a function of the 6.2$\mu$m
  PAH EW, for starbursts (green circles -- Brandl et al. 2006), AGN
  (red squares, Weedman et al. 2006) and ULIRGs (in blue diamonds),
  adapted from Armus et al. (2006). The vertical (red) and horizontal
  (green) lines indicate the fraction of AGN and starburst
  contribution.}
\vspace*{-0.4cm}
\end{figure}

The advent of mid-IR spectroscopy with the Infrared Space Observatory
(ISO) opened a new window in the study of active and starburst
galaxies \cite[see][ for a review]{Verma05}. Emission lines from
relatively high excitation ions which are not produced by O stars can
be used as a direct proble of AGN activity. In the infrared such lines
are the [NeV]$\lambda14.3\mu$m and [NeV]$\lambda24.3\mu$m with an
ionization potential of 97.1eV, and the [OIV]\footnote{The [OIV] line
was easier to detect by ISO, but its predictive power is not as strong
since it can also be produced in areas of strong galactic outflows and
shocks.}$\lambda25.9\mu$m with 54.9eV. The neon lines were detected by
ISO in Seyferts, though their detection in ULIRGs was challenging due
to the limitations in the sensitivity of the ISO spectrographs
\citep{Genzel98, Sturm00}. Another indicator is the strength of the
PAH emission bands which are suppressed in AGN since the hard
radiation field photo-dissociates the PAH molecules \citep{Lutz98}.

The improved sensitivity of IRS on Spitzer \citep{Houck04, Werner04},
a factor of $\sim$100 compared to ISO, enables us to extend these
diagnostics. More detections based on the [NeV] lines were secured for
a large sample of galaxies and an in depth study of PAH emission has
become possible (see Fig. 1). Moreover, techniques employing
correlations based on the strength of other lines such as
[SIV]$\lambda10.5\mu$m and [SIII]$\lambda18.7/33.5\mu$m or pricipal
component analysis are also being explored \citep[see][ for
details]{Weedman05, Armus06, Brandl06, Dale06, Buchanan06}. More
specifically the use of the 6.2$\mu$m PAH is now readily used instead
of the 7.7$\mu$m PAH which is more affected by uncertainties in the
extinction due to the absorption by the 9.7$\mu$m silicate feature.

In addition to the mid-IR high-ionization lines and the 6--18$\mu$m
PAH emission features, the strength of the 3.3$\mu$m PAH has also been
proven a useful AGN tracer. This feature, which can be probed from the
ground for relatively bright sources, has an EW of $\sim$0.1$\mu$m in
starburst systems but is also supressed in AGN
\citep[see][]{Imanishi06}.

\vspace*{-0.3cm}  
\section{Mid-IR Continuum Diagnostics}
\vspace*{-0.3cm}

Even though high ionization line emission is the ideal probe of an
active nucleus, despite the IRS sensitivity, the lines are often
difficult to detect in highly extinct and faint/distant sources. The
presence of an AGN though can also be inferred by detecting the
thermal emission from dust surrounding the putative torus heated by
the accretion disk to nearly sublimation temperatures ($\sim$1000K)
and radiating in near equilibrium. This emission had been detected in
Seyfert galaxies and is now evident in IRS spectra of distant sources
\citep[see][and Fig. 2]{Alonso03}.

\begin{figure}[!ht]
\plotone{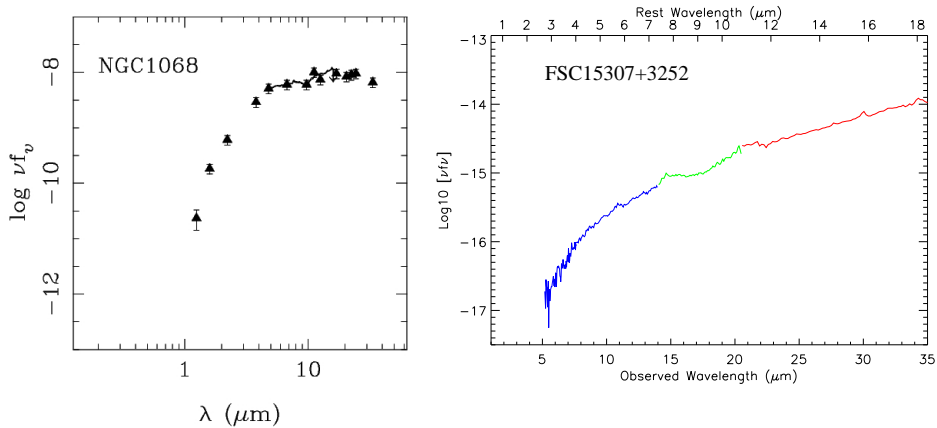}
\caption{The SED of the nucleus of NGC1068, our nearest Sy-2 at
  D=14.4Mpc (left), adapted from Alonso-Herrero et al. (2003) compared
  to the IRS spectrum of FSC15307+3252, a quasar at z=0.92, D=5.8Gpc
  (see Teplitz et al. 2006). Note the similarity in the thermal
  emission from the dust near the AGN torus heated to near sublimation
  temperatures.}
\end{figure}

Unlike extreme starbursts or HII regions which also display a rising
slope at $\lambda>10\mu$m, due to the heating of the grains to
T$\sim$300K by the embedded O/B stars, the presence of this hot
continuum at $\lambda\sim5\mu$m is only seen in AGN. \citet{Laurent00}
proposed an AGN diagnostic method -- now called the ``Laurent
Diagram'' -- which takes advantage of this difference, as well as the
destruction of PAHs in both AGN and extreme starbursts. Using three
template mid-IR spectra for a ``pure'' AGN, an HII region, and a
photodissociation region (PDR) they defined an AGN dominated locus in
a two parameter phase-space (see Fig. 3). The proximity of a galaxy to
one of the three corners, i.e. the AGN corner, would suggest the
extend by which the mid-IR spectrum displays AGN characteristics.

\begin{figure}[!ht]
\plotone{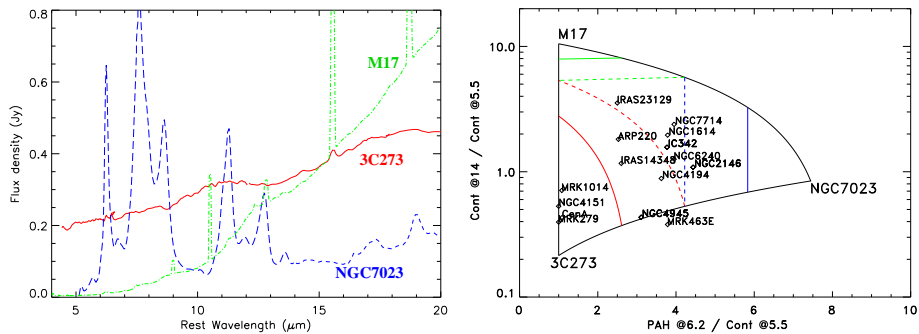}
\caption{Application of the ``Laurent Diagram'' using IRS spectra. On
  the left we display the three template spectra: M17 (scaled by
  10$^{-3}$) as the HII/extreme starburst, 3C273 as the pure AGN
  template, and NGC7023 as the pure PDR template. On the right we show
  the location of a number of starbursts, AGNs and ULIRGs on the
  diagram. The dotted and solid lines indicate the 50 and 75\%
  contribution of the corresponding template to the integrated
  spectrum of a source (see also Armus et al. 2006)}.
\end{figure}
\vspace*{-0.5cm}  

A first application of this method to Spitzer/IRS spectroscopy has
been presented in \citet{Armus06}. In Figure 3 we display the
principle of the method with the three template spectra used, as well
the classification of a number of galaxies based on it. Even though we
suggest the reader to review the original paper for a detailed
discussion of the method, a few points are worth mentioning here.
\begin{itemize}

\item The predictive power of the method depends strongly on the
selection of the template spectra used as the three cornerstones. In
the original paper of \cite{Laurent00} Centaurus A was selected as the
AGN template, while now the wealth of high quality IRS spectra enabled
us to fine tune the selection.

\item The actual coordinates of the three templates depend on the
corresponding PAH to 5.5$\mu$m continuum and 14.5 to 5.5$\mu$m
continuum ratios. Since the continuum fluxes are calculated by
integrating under the spectrum, the wavelength range used for the
integration directly affects the value of the ordinate for all points
in the diagram. It is not obvious that a flat $\nu f_{\nu}$ spectrum
will result in an ordinate equal to 1. In addition, the exact
selection of the limits and underlying continuum in the calculation of
the 6.2$\mu$m PAH, also affects the value of the abscissa.

\item None of the templates have been corrected for extinction, so the
intrinsic extinction in the observed integrated spectrum of a source
may result in placing it outside the parameter space of the
diagram. Note that the nucleus of a galaxy, which may contain a strong
AGN and/or HII region component, is often more enshrouded than the PDR
regions in its disk. As a result of this variable extinction, there is
no one-to-one translation of an extinction vector for all points
placed in the Laurent diagram.

\item The method provides quantitative means for estimating the AGN
  contribution to the ``integrated mid-IR spectrum'' of a galaxy. Even
  though the $\sim25$--$35\mu$m flux traces luminosity it has not yet
  been demonstrated that the this method can be used to estimate the
  contribution of an AGN to the infrared luminosity (L$_{\rm IR}$) of a
  system.

\end{itemize}

The ``Laurent Diagram'' appears to be the most powerful method for AGN
classification in the mid-IR for galaxies where no detections or
strong limits on high ionization lines are available. More analysis
based on Spitzer/IRS spectra will be need to extend its predictive
power to the whole IR. Ongoing work indicates that the method can also
be applied in Spitzer/IRS low resolution spectroscopic studies of
extended star forming regions where variations in the slope of the
mid-IR spectrum and the strength of PAHs can be used to quantify the
intensity of star formation of the regions (Leboutellier priv. comm.).

\section{Theoretical Modeling}
\vspace*{-0.3cm}

Another approach to quantitatively assess the contribution of the
various components in the observed IR emission from a source is to fit
the IR spectrum using theoretical models. Ideally one could use a full
3D radiative transfer calculation to model the dust properties and
match the observed spectrophotometry. A large set of theoretical SEDs
are now available for comparison with the observations
\citep[i.e.][and references therein.]{Elitzur06, Ralf06}. However,
limited knowledge of the details in the spatial distribution and
geometry of the sources, in particular for systems such as starburst
and ULIRGs, which are inherently disturbed, often make this approach
rather challenging.

Recently, a new fitting approach relying on the Spitzer/IRS
5--38$\mu$m spectra, using constrains from near- and far-IR
observations has been developed \citep{Marshall06}. The method assumes
that the SED of a galaxy can be decomposed into dust components at
different characteristic temperatures, source emission components,
embedded photospheric emission from starburst cores or an active
nucleus as well as PAH feature emission. The emission from each
component is calculated using a realistic dust model consisting of a
distribution of thermally emitting carbonaceous and silicate
grains. The model accounts for stochastic emission from very small
grains by fitting model PAH templates to the spectra. This method has
been applied in a variety of Spitzer/IRS data and most sources are
well fit by a maximum of four components referred as cold ($\sim$30K),
tepid ($\sim$100K), warm ($\sim$200K), and hot ($>$300K). The results
of the fits are used to calculate the relative bolometric
contributions from the different components, providing a method to
compare the infrared properties of starburst galaxies and
quasars. Sample results of the fitting method on NGC7714 -- a
starburst galaxy -- a ULIRG such as NGC6240 and the quasar PG0804+761
are presented in Figure 4.

\vspace*{-0.2cm} 
\begin{figure}[!ht]
\plotone{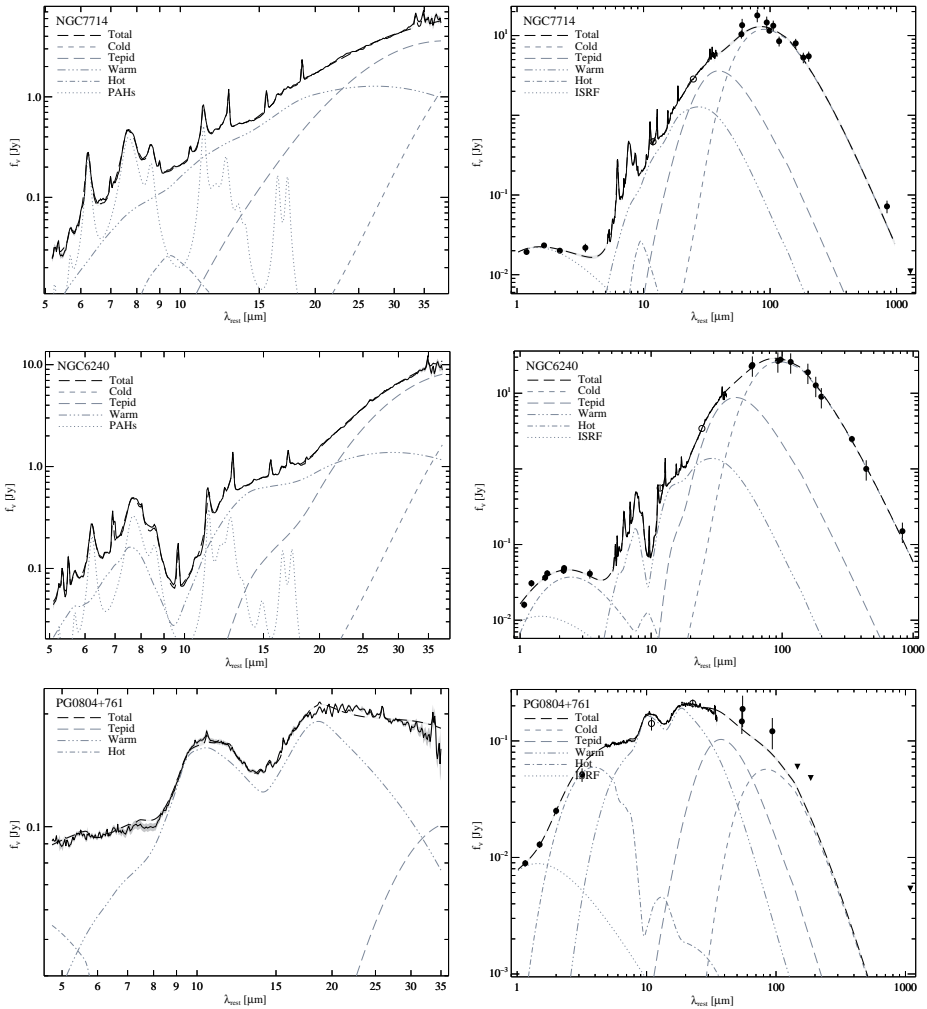}
\vspace*{-0.3cm} 
\caption{Model fits to the IRS spectrum and SED of NGC7714, NGC6240,
  and PG0804+761. The contribution of the various temperature
  components is indicated by the dotted and dashed lines. Note how
  well the model fit trace, PAH emission features, as well as the
  silicate absorption and -- in the case of PG0804+761 -- emission
  features. The presence of an AGN in PG0804+761 is also revealed by
  the hot dust component peaking at $\sim4\mu$m. For more details on
  the luminosity contribution of each component see Marshall et
  al. (2006)}
\end{figure}

\vspace*{-0.3cm} 
\section{Conclusions}
\vspace*{-0.3cm} 

It has become evident from the numerous contributions presented during
this meeting that the high quality of Spitzer/IRS spectroscopy is
opening new horizons in the use of the infrared as a tracer of the
properties of nearby and high-redshift galaxies. With more than three
years of mission to go, we have just glimpsed on the possibilities
that lie ahead.

\acknowledgements 

I would like to thank Jim Houck for giving me the opportunity to spend
5.5 formative years in Ithaca and participate in the Spitzer/ IRS
``venture''. I would also like to acknowledge all members of the IRS
instrument team at Cornell and Caltech for many enlightening and
stimulating discussions on spectroscopy and the infrared, as well as
the organizers for the invitation to this exciting conference. I also
appreciate the help of Lee Armus and Jason Marshall who provided data
and plots for this paper prior to publication.

\end{document}